    \pdfoutput=1
    
    \documentclass[11pt]{article}
    
    \usepackage[]{ACL2023}
    
    \usepackage{times}
    \usepackage{latexsym}
    \usepackage{graphicx}
    \usepackage{xurl}
    \usepackage{microtype}

    \usepackage[T1]{fontenc}
    
    \usepackage[utf8]{inputenc}
    
    \usepackage{microtype}
    
    \usepackage{inconsolata}
    \usepackage{booktabs}
    
    \title{Toward Low-Latency End-to-End Voice Agents for Telecommunications Using Streaming ASR, Quantized LLMs, and Real-Time TTS}
    
\author{
  Vignesh Ethiraj\thanks{\hspace{0.5em}These authors contributed equally to this work.} \quad
  Ashwath David\footnotemark[1] \quad
  Sidhanth Menon\footnotemark[1] \quad
  Divya Vijay\footnotemark[1] \\
  NetoAI \\
  \texttt{support@netoai.ai}
}

\begin{document}
    \maketitle 
    
    \begin{abstract}
    
We propose a low-latency, end-to-end voice-to-voice communication pipeline, purpose-built for real-time, interactive telecom scenarios such as call center automation and conversational IVR (Interactive Voice Response) systems. Our system integrates streaming automatic speech recognition (ASR), a 4-bit quantized large language model (LLM), retrieval-augmented generation (RAG) over telecom documents, and real-time text-to-speech (TTS) to enable responsive, knowledge-grounded spoken interactions. To evaluate performance in a realistic setting, we constructed a custom dataset of 500 human-recorded utterances featuring telecommunications-related questions sourced from RFC documents. This benchmark simulates user queries to a telecom voice agent and supports analysis of both latency and domain relevance. The pipeline combines sentence-level streaming, concurrent processing, and vector-based document retrieval, achieving real-time factors (RTF) below 1.0 across components. Results demonstrate the system's effectiveness for low-latency telecom applications such as customer support and diagnostics.

    \end{abstract}
    
    \section{Introduction}
    
    Real-time speech interfaces increasingly demand low-latency processing across ASR, Natural Language Understanding (NLU), and TTS. While significant advances have been made in these areas individually, integrating them into a single, low-latency, end-to-end system remains challenging. Naively chaining models sequentially results in cumulative delays, limiting practical usability in conversational agents, voice summarization systems, and assistive technologies.
    
    This work addresses the challenge of minimizing end-to-end latency in a speech-input-to-speech-output pipeline. We combine a pre-trained ASR model, a quantized LLM for summarization, and a real-time TTS system, connected via a multi-threaded streaming architecture. Our contributions include:
    
\begin{itemize}
  \item \textbf{Sentence-level streaming:} 
    The pipeline supports sentence-level streaming, allowing the LLM to transmit generated sentences incrementally to the TTS module for early and continuous audio output.

  \item \textbf{4-bit LLM quantization:} 
    The LLM is quantized to 4-bit precision, significantly reducing GPU memory footprint and inference latency while preserving generation quality.

  \item \textbf{Concurrent module execution:} ASR, LLM, and TTS modules operate concurrently. They are coordinated via a non-blocking producer-consumer pattern, enabling seamless real-time processing without blocking or unnecessary wait times.

  \item \textbf{Latency and performance analysis:} 
    The system includes a detailed breakdown of latency components, enabling understanding of performance trade-offs and the effects of each module and architectural choice on overall responsiveness.
\end{itemize}

    \section{Related Work}
    
    \begin{figure*}[t]
        \centering
        \includegraphics[width=\textwidth]{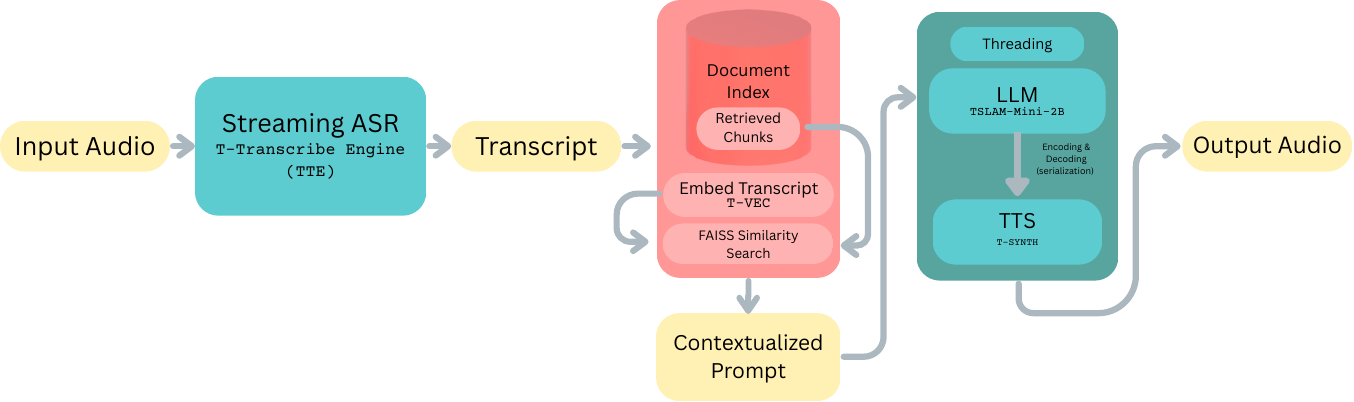}
        \caption{Overall pipeline architecture.}
        \label{fig:pipeline_architecture}
    \end{figure*}
    
    Recent advances in streaming ASR, quantized LLMs, and real-time TTS have enabled the development of low-latency voice transformation systems through careful architectural choices and optimization techniques that directly inform our pipeline implementation.

\subsection{Conformer-Based Streaming ASR Models}

Conformer-based ASR models have become the main approach for streaming speech recognition. They effectively capture both local and global acoustic dependencies using convolutional and self-attention layers~\cite{gulati2020conformer}. 

NVIDIA's NeMo framework~\cite{nemoasr2025} offers optimized, pre-trained Conformer models such as \texttt{nvidia/stt\_en\_conformer\_ctc\_small}. These achieve real-time factors below 0.2 and competitive word error rates (WER) on benchmarks like LibriSpeech~\cite{panayotov2015librispeech}. 

Such models use con\-nec\-tion\-ist tem\-po\-ral clas\-si\-fi\-ca\-tion (CTC) for align\-ment-free, frame-syn\-chro\-nous output.

Other state-of-the-art toolkits and models, such as AI4Bharat's IndicConformer for multilingual Indian ASR~\cite{ai4bharat2024indic}, AssemblyAI's Conformer-1 model~\cite{assemblyai2023conformer1}, and the open-source SpeechBrain toolkit~\cite{speechbrainconformer2022}, further expand Conformer adoption to large-scale multilingual, noisy, and domain-specific speech tasks. Our pipeline leverages our proprietary, telecom-optimized \texttt{T-Transcribe Engine (TTE)}, based on a Conformer-CTC architecture. Designed specifically for real-time conversational and call-center scenarios, TTE balances recognition accuracy with low-latency inference to meet the stringent requirements of telecom domain applications.

    \subsection{4-bit Quantization for LLM Deployment}
    Post-train\-ing quan\-ti\-za\-tion tech\-niques have proven ef\-fec\-tive for de\-ploy\-ing large lan\-guage mod\-els in resource-con\-strained en\-vi\-ron\-ments. 
    Using the \texttt{BitsAndBytesConfig} frame\-work enables 4-bit quan\-ti\-za\-tion with min\-i\-mal per\-for\-mance degra\-da\-tion, achiev\-ing up to 40\% la\-ten\-cy re\-duc\-tion while pre\-serv\-ing gen\-er\-a\-tion qual\-i\-ty \cite{Gong2022QuantLLM}. Re\-cent work on quan\-tized con\-ver\-sa\-tion\-al mod\-els demon\-strates that 4-bit pre\-ci\-sion main\-tains over 95\% of orig\-i\-nal per\-for\-mance while re\-duc\-ing com\-pu\-ta\-tion\-al com\-plex\-i\-ty by fac\-tors of 60× or more \cite{Chen2023QUADS}.
    
    \subsection{Streaming TTS and Parallel Synthesis}
     Modern neural vocoders have achieved real-time synthesis through tensor-level optimizations and chunked processing frameworks. Streaming TTS systems that interleave text encoding and waveform generation can reduce time-to-first-audio to under 50ms \cite{Ellinas_2020, Lee2023MetaVoc}.
    
\subsection{RAG Integration in Voice Systems}

Retrieval-Augmented Generation (RAG) architectures combine neural retrievers, such as dense embedding models trained for semantic search with generative language models, allowing new information to be dynamically included in system responses without retraining the core model~\cite{lewis2020rag}. Recent work demonstrates that RAG techniques have been successfully extended from text-only settings to voice and multimodal systems:

\begin{itemize}
  \item WavRAG is a pioneering audio-integrated RAG framework. It enables spoken dialogue models to retrieve and utilize both audio and text knowledge bases, with end-to-end audio support for real-time, context-aware conversation~\cite{galhotra2025wavrag}.
  \item VoxRAG further enhances this approach by implementing transcription-free, speech-to-speech retrieval. This allows query and answer generation entirely in the audio domain, showcasing the feasibility of RAG for real-world, spoken question answering tasks~\cite{rackauckas2025voxrag}.
  \item RAG-based agents are already deployed in voice assistants and IVR systems. These combine speech-to-text, neural document retrieval, generative language models, and text-to-speech synthesis for accurate, context-rich spoken interactions, especially in customer service and enterprise settings~\cite{sambare2025voice}.
\end{itemize}

These developments have made RAG a foundational technique for enriching voice systems with up-to-date, domain-specific information at inference time.

    \section{Pipeline Architecture and Implementation}
    
    Our end-to-end voice transformation pipeline integrates streaming ASR, RAG, quantized LLM inference, and real-time TTS synthesis using a modular multi-threaded architecture designed to minimize end-to-end latency.
    
    \subsection{Streaming ASR Module}
    
    We employ our \texttt{T-Transcribe Engine (TTE)} model, a Conformer-based architecture optimized for real-time speech recognition with Connectionist Temporal Classification (CTC) training. This model effectively balances low latency and transcription accuracy with sub-0.2 real-time factor (RTF) on GPU, making it suitable for streaming applications. The ASR module transcribes audio waveforms loaded via \texttt{soundfile} and generates input text transcripts with precise timings recorded for downstream metrics.
    
    \subsection{Retrieval-Augmented Generation (RAG)}
    
    To enhance factual grounding and contextual relevance of the LLM responses, we integrate a RAG submodule based on FAISS~\cite{faisslib2024} similarity search over document embeddings. Document indexing is performed offline or at startup using the \texttt{NetoAISolutions/T-VEC} model \cite{tvec2025retrieval}, which encodes documents into normalized dense vectors. When a serialized FAISS index and corresponding documents are cached, these are loaded for efficiency; otherwise, the system builds the index from text files in a configurable directory. Retrieval queries embed the ASR transcript and perform inner-product search with configurable \texttt{k} neighbors to provide relevant context documents concatenated as input to the LLM prompt. This design leverages efficient nearest neighbor search algorithms to maintain sub-second retrieval latency and integrates seamlessly with the generation stage.
    
    \subsection{Quantized Large Language Model (LLM) Inference}
    
    We utilize the \texttt{NetoAISolutions/TSLAM-Mini-2B} \cite{tslam2025retrieval} causal LLM, loaded via Hugging Face Transformers~\cite{wolf2020transformers}, applying 4-bit post-training quantization via the \texttt{BitsAndBytes} ~\cite{dettmers2022llmint8} library to reduce GPU memory footprint and enable faster inference without significant quality loss~\cite{Gong2022QuantLLM}. The tokenizer is configured with padding tokens to support variable-length input sequences safely. Streaming generation is implemented with a custom \texttt{PunctuatedBufferStreamer} class that segments output text into sentences in real-time using regex-based punctuation detection and places serialized sentences in a thread-safe queue. This streamer also captures fine-grained latency metrics such as the time-to-first-token generation.
    
    \subsection{Real-Time Text-to-Speech Pipeline}
    
    The TTS submodule is implemented via our proprietary, telecom-optimized \texttt{T-SYNTH} TTS model, which leverages a vocal synthesis pipeline initialized with a warmup routine on a reference voice to reduce latency jitter by preloading required components. The synthesis is performed in a dedicated thread that consumes sentences serialized by the streamer and converts text to waveform chunks. These audio chunks are asynchronously post-processed and concatenated to produce a single WAV output file. Sentence-level synthesis timings are recorded, allowing for detailed analysis of TTS synthesis overhead.

    \subsection{Multi-threading and Synchronization}
    
    Our implementation uses threading to parallelize LLM generation and TTS synthesis, with custom sentence streaming that segments LLM output in real-time and feeds it to the TTS pipeline, achieving sub-second response times. The sentence queue was given a timeout of 0.05 seconds to receive each chunk of the LLM response as it came in, with elements in the queue being separated by appropriate punctuations by the \texttt{PunctuatedBufferStreamer}. The TTS thread was made to begin before the LLM thread to let it load the necessary components before the first LLM response tokens started coming in. A warm-up dummy TTS pipeline was also introduced for the same purpose, as highlighted previously. A technique of binary serialization was also introduced between the LLM response and the TTS generation to further reduce our pipeline time. The response was packed into binary serials, and it was then unpacked at the time of TTS generation. \texttt{msgpack} was used to achieve this. These methods helped reduce the end-to-end pipeline time significantly, by about ~0.8-1.0 seconds.

    \subsection{Metrics Reporting and Performance Profiling}
    
    Performance and resource utilization metrics are diligently tracked throughout the pipeline. The \texttt{MetricsReporter} class collates timers for model loading, ASR processing, RAG retrieval, LLM generation, and TTS synthesis. It reports latency breakdowns including ASR speed in words/sec, real-time factors, LLM tokens/sec, and time-to-first-audio for user experience insights. GPU details and memory consumption are also included, enabling system-level profiling.
     
The foundation of our low-latency voice transformation pipeline is built upon a combination of the following techniques:
\begin{itemize}
    \item Conformer-based streaming ASR
    \item 4-bit quantized LLMs
    \item Parallel LLM response and TTS synthesis
    \item Binary serialization
    \item Efficient RAG retrieval
    \item Optimized threading
\end{itemize}

    \subsection{Model Initialization}
    
    Models are loaded once, and quantized models are initialized using \texttt{AutoModelForCausalLM} with \texttt{BitsAndBytesConfig}. TTS is warmed up with a dummy sentence to reduce first-inference latency.
    
    \subsection{Streaming Sentence Generation}
    
    We introduce a \urlstyle{tt}\url{PunctuatedBufferStreamer} class based on the HuggingFace {\urlstyle{tt}\url{TextStreamer}}. It detects full sentences using punctuation and pushes them into a thread-safe queue. This allows the TTS system to begin audio synthesis while LLM generation is ongoing.
    
    \subsection{Multi-threaded Execution}
    
    LLM generation and TTS decoding run in parallel threads. The LLM acts as a producer, and the TTS system consumes sentences in FIFO order. ASR is processed upfront due to its short execution time.

    \begin{table*}[t]
    \centering
    \label{tab:metrics_summary}
    \resizebox{\textwidth}{!}{%
    \begin{tabular}{lcccccccccc}
    \toprule
    \textbf{Stat} & \textbf{ASR} & \textbf{RAG} & \textbf{LLM} & \textbf{TTS} & \textbf{Total} & \textbf{ASR Speed}  & \textbf{LLM Speed} & \textbf{Cosine} & \textbf{TTFT} & \textbf{TTFA} \\
                  & \textbf{Processing}         & \textbf{Retrieval}       & \textbf{Generation}         & \textbf{Synthesis}       & \textbf{Time}               & \textbf{(words/sec)} 
                  & \textbf{(tokens/sec)}          & \textbf{Similarity (\%)}  \\
    \midrule
    \textbf{Mean} & 0.049 & 0.008 & 0.670 & 0.286 & 0.934 & 394.18 & 80.06 & 0.873 & 0.106 & 0.678 \\
    \textbf{Min}  & 0.029 & 0.008 & 0.218 & 0.106 & 0.417 & 134.24 & 58.60 & 0.659 & 0.077 & 0.412 \\
    \textbf{Max}  & 0.069 & 0.012 & 1.706 & 1.769 & 3.154 & 1010.15 & 86.97 & 1.000 & 0.181 & 1.482 \\
    \bottomrule
    \end{tabular}%
    }
    \caption{Latency and performance metrics across pipeline components. All times in seconds except where mentioned.}
    \end{table*}

    \bigskip
    
    \begin{figure*}[t]
        \centering
        \includegraphics[width=\textwidth]{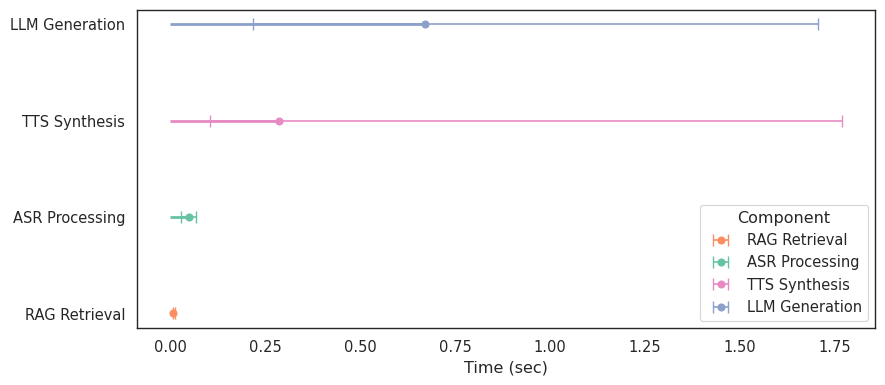}
        \caption{Average latencies with min-max error bars }
        \label{fig:latency_bars}
    \end{figure*}
    
    \section{Experimental Setup}
    \subsection{Dataset}

    To evaluate our telecom-oriented voice-to-voice system, we constructed a custom dataset of 500 human-recorded utterances, each corresponding to a spoken telecommunications-related question. The prompts were sourced from open-access RFC (Request for Comments) documents to ensure content relevance and compatibility with the downstream RAG-based retrieval system used by our agent.
    
    While large-scale voice datasets such as LibriSpeech \cite{panayotov2015librispeech}, Common Voice \cite{ardila2020common}, and the Spoken Wikipedia Corpora \cite{kohn2018spokenwiki} offer broad linguistic coverage, they are either domain-agnostic or narratively structured, making them suboptimal for evaluating question-driven, real-time dialog systems in specialized verticals like telecommunications. Moreover, these corpora do not offer fine-grained alignment with structured backend knowledge (e.g., RFC indices), a critical feature for retrieval-augmented generation (RAG) pipelines.
    
    Our dataset bridges this gap by aligning natural speech inputs with a vector index derived from the same RFC documents. This alignment enables controlled and realistic simulation of query-response behavior in telecom use cases. Utterances were recorded by two different speakers with varied accents and speaking rates to approximate deployment variability. Each file is stored in WAV format and was manually verified for clarity and timestamp consistency. The dataset totals approximately 45 minutes of audio, with an average utterance duration of 6.36 seconds.
    
    This design supports robust benchmarking of streaming latency, transcription accuracy, and sentence-level retrieval performance in a setting that closely mirrors the demands of real-world telecom voice agents.

    \subsection{Hardware}
    All experiments were conducted on a system equipped with an NVIDIA H100 GPU (80GB)and 256 GB of RAM. The H100’s high memory bandwidth and transformer-optimized architecture enabled real-time inference for quantized LLMs, streaming ASR, and TTS components. All software components were executed in a single-node setup using mixed-precision inference where supported.
    
    \section{Evaluation and Results}
    
    \subsection{Real-Time Performance}
    
    Our pipeline achieves near real-time performance across all components. The average total latency per utterance is \textbf{0.94 seconds}, comfortably under the 1-second threshold typically considered acceptable for interactive systems. The \textbf{ASR} and \textbf{TTS} modules operate with mean latencies of around \textbf{0.05s} and \textbf{0.28s}, respectively, while the \textbf{LLM} is expectedly the most time-consuming, averaging \textbf{0.67s} per generation. Retrieval latency is negligible on average (\textbf{0.008s}).
    
    \subsection{Streaming Efficiency}
    
    To assess how the system handles input/output streaming, we report \textbf{time-to-first-token (TTFT)} and \textbf{time-to-first-audio (TTFA)}. The average TTFT is \textbf{0.106s}, suggesting that text generation begins quickly after receiving input. TTFA is slightly higher at \textbf{0.678}, which includes LLM latency and the initial TTS processing. This shows that there is an average gap of around 0.5 seconds between the LLM generating its first set of tokens and the TTS beginning its synthesis. 
    
    \subsection{Model Throughput and Speed}
    
    The ASR operates at an average of \textbf{394 tokens/sec}, and the LLM generates tokens at an average of \textbf{80 tokens/sec}. These speeds indicate that the system is well-suited for real-time applications, with ample headroom for longer utterances or simultaneous streams.
    
    \subsection{Semantic Preservation}
    
    Cosine similarity between the ASR transcript embeddings and the LLM-generated outputs averages \textbf{0.87}, indicating strong semantic preservation during transformation. 
    
    \subsection{Latency Variability}
    
    While mean latencies are low, the worst-case total pipeline latency reached \textbf{3.154 seconds}. Given that the best case was at a low latency \textbf{0.417 seconds}, with the average being less than 1, it can be inferred that GPU processing fluctuations were at play in determining the average pipeline latency. These outliers of 2+ seconds are rare, and this is further supported by our general observations of the pipeline usually settling comfortably at less than \textbf{0.7 seconds} in most test runs. 
    
    \section{Conclusion}
    
    We have introduced a low-latency, end-to-end voice agent pipeline tailored for telecommunications applications, integrating streaming ASR, retrieval-augmented generation with a quantized LLM, and real-time TTS synthesis using a modular, multi-threaded framework. Our detailed evaluation demonstrates that streaming workflows, sentence-level concurrency, aggressive quantization, and efficient RAG significantly reduce total system latency while preserving response quality and semantic relevance. With an average response time below 1s and strong performance across all components, the pipeline meets the demanding requirements of real-time interactive voice scenarios such as customer support, diagnostics, and IVR replacement.
    
    By open-sourcing our dataset and providing a reproducible methodology, we lay the groundwork for future research on scalable, low-latency spoken dialog systems. We believe the techniques outlined - spanning ASR architectures, LLM quantization, document retrieval, and producer-consumer parallelism - can be adapted to other verticals where fast, knowledge-augmented voice interfaces are essential. Future work will address scaling to more diverse domains, supporting multilingual use cases, and integrating adaptive learning for continual improvement in real-world deployments.

    \section*{Limitations}
    A key limitation of our pipeline lies in the automatic speech recognition (ASR) output used for downstream processing. Inaccuracies in transcription, especially in cases involving abbreviations, proper nouns, or domain-specific terms can lead to degraded performance in similarity computations in RAG. Since our cosine similarity metric is computed over these transcriptions, ASR errors directly impact the final similarity scores. 
    
    Future work should consider adopting larger or more specialized ASR models that are even better adapted to the task domain. Improved ASR performance would help mitigate transcription errors and thereby enhance the quality of semantic similarity assessments.
    
    These enhancements have the potential to improve robustness and better capture the intended meaning in noisy, imperfect, or domain-specific transcriptions.

    \section*{Ethics Statement}
    
    There are no ethical concerns to be discussed in this implementation. 
    
    \bibliography{custom}

\begin{thebibliography}{21}
\expandafter\ifx\csname natexlab\endcsname\relax\def\natexlab#1{#1}\fi

\bibitem[{AI4Bharat(2024)}]{ai4bharat2024indic}
AI4Bharat. 2024.
\newblock Indicconformer: Multilingual conformer asr models.
\newblock \url{https://huggingface.co/ai4bharat/indic-conformer-600m-multilingual}.

\bibitem[{Ardila et~al.(2020)Ardila, Branson, Davis, Kohler, Meyer, Henretty, Morais, Saunders, Tyers, and Weber}]{ardila2020common}
Rosana Ardila, Megan Branson, Kelly Davis, Michael Kohler, Josh Meyer, Michael Henretty, Reuben Morais, Lindsay Saunders, Francis~M. Tyers, and Gregor Weber. 2020.
\newblock \href {https://arxiv.org/abs/1912.06670} {{Common Voice}: A massively-multilingual speech corpus}.
\newblock \emph{arXiv preprint arXiv:1912.06670}.

\bibitem[{AssemblyAI(2023)}]{assemblyai2023conformer1}
AssemblyAI. 2023.
\newblock Conformer-1: A robust speech recognition model trained on 1 million hours of audio.
\newblock \url{https://assemblyai.com/blog/conformer-1}.

\bibitem[{Biswas et~al.(2025)Biswas, Khan, and Islam}]{Chen2023QUADS}
Subrata Biswas, Mohammad Nur~Hossain Khan, and Bashima Islam. 2025.
\newblock \href {https://arxiv.org/abs/2505.14723} {Quads: Quantized distillation framework for efficient speech language understanding}.
\newblock \emph{arXiv preprint arXiv:2505.14723}.
\newblock Accepted at INTERSPEECH 2025.

\bibitem[{Chen et~al.(2025)Chen, Ji, Wang, Wang, Chen, He, Xu, and Zhao}]{galhotra2025wavrag}
Yifu Chen, Shengpeng Ji, Haoxiao Wang, Ziqing Wang, Siyu Chen, Jinzheng He, Jin Xu, and Zhou Zhao. 2025.
\newblock \href {https://arxiv.org/abs/2502.14727} {Wavrag: Audio-integrated retrieval augmented generation for spoken dialogue models}.
\newblock In \emph{Proceedings of the 2025 Conference on Empirical Methods in Natural Language Processing}.

\bibitem[{Dettmers et~al.(2022)Dettmers, Lewis, Belkada, and Zettlemoyer}]{dettmers2022llmint8}
Tim Dettmers, Mike Lewis, Younes Belkada, and Luke Zettlemoyer. 2022.
\newblock \href {https://arxiv.org/abs/2208.07339} {Llm.int8(): 8-bit matrix multiplication for transformers at scale}.
\newblock \emph{arXiv preprint arXiv:2208.07339}.

\bibitem[{Douze et~al.(2024)Douze, Guzhva, Deng, Johnson, Szilvasy, Mazaré, Lomeli, Hosseini, and Jégou}]{faisslib2024}
Matthijs Douze, Alexandr Guzhva, Chengqi Deng, Jeff Johnson, Gergely Szilvasy, Pierre-Emmanuel Mazaré, Maria Lomeli, Lucas Hosseini, and Hervé Jégou. 2024.
\newblock \href {https://arxiv.org/abs/2401.08281} {The faiss library}.
\newblock \emph{arXiv preprint arXiv:2401.08281}.

\bibitem[{Ellinas et~al.(2020)Ellinas, Vamvoukakis, Markopoulos, Chalamandaris, Maniati, Kakoulidis, Raptis, Sung, Park, and Tsiakoulis}]{Ellinas_2020}
Nikolaos Ellinas, Georgios Vamvoukakis, Konstantinos Markopoulos, Aimilios Chalamandaris, Georgia Maniati, Panos Kakoulidis, Spyros Raptis, June~Sig Sung, Hyoungmin Park, and Pirros Tsiakoulis. 2020.
\newblock \href {https://doi.org/10.21437/interspeech.2020-2464} {High quality streaming speech synthesis with low, sentence-length-independent latency}.
\newblock In \emph{Interspeech 2020}, interspeech\_2020, pages 2022--2026. ISCA.

\bibitem[{Ethiraj et~al.(2025{\natexlab{a}})Ethiraj, Menon, and Vijay}]{tvec2025retrieval}
Vignesh Ethiraj, Sidhanth Menon, and Divya Vijay. 2025{\natexlab{a}}.
\newblock \href {http://arxiv.org/abs/2504.16460} {T-vec: A telecom-specific vectorization model with enhanced semantic understanding via deep triplet loss fine-tuning}.

\bibitem[{Ethiraj et~al.(2025{\natexlab{b}})Ethiraj, Vijay, Menon, and Berscilla}]{tslam2025retrieval}
Vignesh Ethiraj, Divya Vijay, Sidhanth Menon, and Heblin Berscilla. 2025{\natexlab{b}}.
\newblock \href {http://arxiv.org/abs/2505.07877} {Efficient telecom specific llm: Tslam-mini with qlora and digital twin data}.

\bibitem[{Gong et~al.(2023)Gong, Liu, Wang, Yang, Wang, Wu, Xian, Zhao, and Yan}]{Gong2022QuantLLM}
Zhuocheng Gong, Jiahao Liu, Qifan Wang, Yang Yang, Jingang Wang, Wei Wu, Yunsen Xian, Dongyan Zhao, and Rui Yan. 2023.
\newblock \href {https://aclanthology.org/2023.findings-acl.511/} {Prequant: A task-agnostic quantization approach for pre-trained language models}.
\newblock In \emph{Findings of the Association for Computational Linguistics: ACL 2023}, pages 8065--8079.

\bibitem[{Gulati et~al.(2020)Gulati, Qin, Chiu, Parmar, Zhang, Yu, Han, Wang, Zhang, Wu, and Pang}]{gulati2020conformer}
Anmol Gulati, James Qin, Chung-Cheng Chiu, Niki Parmar, Yu~Zhang, Jiahui Yu, Wei Han, Shibo Wang, Zhengdong Zhang, Yonghui Wu, and Ruoming Pang. 2020.
\newblock \href {https://arxiv.org/abs/2005.08100} {Conformer: Convolution-augmented transformer for speech recognition}.
\newblock In \emph{Proc. Interspeech}.

\bibitem[{K{\"o}hn et~al.(2016)K{\"o}hn, Stegen, and Baumann}]{kohn2018spokenwiki}
Arne K{\"o}hn, Florian Stegen, and Timo Baumann. 2016.
\newblock \href {https://aclanthology.org/L16-1735/} {Mining the spoken {W}ikipedia for speech data and beyond}.
\newblock In \emph{Proceedings of the Tenth International Conference on Language Resources and Evaluation ({LREC}'16)}, pages 4644--4647, Portoro{\v{z}}, Slovenia. European Language Resources Association (ELRA).

\bibitem[{Lee et~al.(2023)Lee, Ping, Ginsburg, and Catanzaro}]{Lee2023MetaVoc}
Jungil Lee, Wei Ping, Boris Ginsburg, and Bryan Catanzaro. 2023.
\newblock \href {https://doi.org/10.1109/ICASSP49357.2023.10098974} {Metavoc: Meta-learning for few-shot text-to-speech with optimal transport}.
\newblock In \emph{Proc. IEEE International Conference on Acoustics, Speech and Signal Processing (ICASSP)}, pages 1--5.

\bibitem[{Lewis et~al.(2020)Lewis, Perez, Piktus, Petroni, Karpukhin, Goyal, Kuttler, Lewis, tau Yih, Rockt\"aschel, Riedel, and Kiela}]{lewis2020rag}
Patrick Lewis, Ethan Perez, Aleksandra Piktus, Fabio Petroni, Vladimir Karpukhin, Naman Goyal, Heinrich Kuttler, Mike Lewis, Wen tau Yih, Tim Rockt\"aschel, Sebastian Riedel, and Douwe Kiela. 2020.
\newblock Retrieval-augmented generation for knowledge-intensive nlp tasks.
\newblock In \emph{Advances in Neural Information Processing Systems}, volume~33, pages 9459--9474.

\bibitem[{NVIDIA(2025)}]{nemoasr2025}
NVIDIA. 2025.
\newblock Models: Nemo asr collection.
\newblock \url{https://docs.nvidia.com/nemo-framework/user-guide/latest/nemotoolkit/asr/models.html}.

\bibitem[{Panayotov et~al.(2015)Panayotov, Chen, Povey, and Khudanpur}]{panayotov2015librispeech}
Vassil Panayotov, Guoguo Chen, Daniel Povey, and Sanjeev Khudanpur. 2015.
\newblock \href {https://doi.org/10.1109/ICASSP.2015.7178964} {Librispeech: an asr corpus based on public domain audio books}.
\newblock In \emph{2015 IEEE International Conference on Acoustics, Speech and Signal Processing (ICASSP)}, pages 5206--5210. IEEE.

\bibitem[{Rackauckas and Hirschberg(2025)}]{rackauckas2025voxrag}
Zackary Rackauckas and Julia Hirschberg. 2025.
\newblock \href {https://arxiv.org/abs/2505.17326} {Voxrag: A step toward transcription-free rag systems in spoken question answering}.
\newblock \emph{arXiv preprint arXiv:2505.17326}.

\bibitem[{Sambare et~al.(2025)Sambare, Kadam, Agre, Chandravanshi, Agrawal, and Samaga}]{sambare2025voice}
G.~B. Sambare, Ganesh Kadam, Aditya Agre, Amay Chandravanshi, Kanak Agrawal, and Parinitha Samaga. 2025.
\newblock \href {https://www.iosrjournals.org/iosr-jce/papers/Vol27-issue2/Ser-3/D2702032640.pdf} {Advancements in voice-activated systems: A comprehensive survey on retrieval-augmented generation (rag) and large language model techniques}.
\newblock \emph{IOSR Journal of Computer Engineering}, 27(2):26--40.

\bibitem[{SpeechBrain(2022)}]{speechbrainconformer2022}
SpeechBrain. 2022.
\newblock Streaming speech recognition with conformers.
\newblock \url{https://speechbrain.readthedocs.io/en/v1.0.2/tutorials/nn/conformer-streaming-asr.html}.

\bibitem[{Wolf et~al.(2020)Wolf, Debut, Sanh, Chaumond, Delangue, Moi, Cistac, Rault, Louf, Funtowicz, Davison, Shleifer, von Platen, Ma, Jernite, Plu, Xu, Scao, Gugger, Drame, Lhoest, and Rush}]{wolf2020transformers}
Thomas Wolf, Lysandre Debut, Victor Sanh, Julien Chaumond, Clement Delangue, Anthony Moi, Pierric Cistac, Tim Rault, R{\'e}mi Louf, Morgan Funtowicz, Joe Davison, Sam Shleifer, Patrick von Platen, Clara Ma, Yacine Jernite, Julien Plu, Canwen Xu, Teven~Le Scao, Sylvain Gugger, Mariama Drame, Quentin Lhoest, and Alexander~M. Rush. 2020.
\newblock \href {https://www.aclweb.org/anthology/2020.emnlp-demos.6} {Transformers: State-of-the-art natural language processing}.
\newblock In \emph{Proceedings of the 2020 Conference on Empirical Methods in Natural Language Processing: System Demonstrations}, pages 38--45, Online. Association for Computational Linguistics.

\end{thebibliography}
    \bibliographystyle{acl_natbib}

    \end{document}